\documentstyle[eqsecnum,aps]{revtex}

\newcommand{\bml}{\begin{mathletters}}
\newcommand{\eml}{\end{mathletters}}
\newcommand{\bey}{\begin{eqnarray}}
\newcommand{\eey}{\end{eqnarray}}
\newcommand{\be}{\begin{equation}}
\newcommand{\ee}{\end{equation}}

\begin{document}

\title{The Light-Front SU(N) Yang-Mills Theory for the
Weyl gauge }

\author{Jerzy A. Przeszowski\thanks{email: jprzeszo@ippt.gov.pl}}
\address{Institute of Fundamental Technological Research\\
Polish Academy of Sciences\\
ul. \'Swi\c{e}tokrzyska 11/21, \ 00-049 Warsaw, \ Poland}
\date{\today}

\maketitle
\begin{abstract}
The canonical quantization is performed at a light-front surface
for the SU(N) Yang-Mills theory. The Weyl gauge is imposed as a
gauge condition. The suitable parameterization is
chosen for the transverse gauge field components in order to have
Dirac brackets independent of interactions. The generating
functional is defined for the perturbation theory and it is shown
to coincide with its equal-time counterpart.
\end{abstract}

\section{Introduction}
The light-front (LF) quantization (at a fixed $x^{+}$ surfaces)
\cite{Dirac} of the gauge field models is almost exclusively
performed for the light-cone (LC) gauge condition: 
$A_{-}^a = A^{a+} = 0$ \cite{BrodPaulPin1998}. Though this choice
leads to a considerable simplifications of the interaction
Hamiltonians, then it leads directly to the Cauchy Principal Value
(CPV) prescription for the spurious poles of the perturbative gauge
field propagators - which is known to be an inconsistent
prescription for the higher order computations
\cite{BassettoNardeliSoldati1991}. The consistent regularization is
the causal Mandelstam-Leibbrandt (ML) prescription \cite{ML}, which
has been first obtained as a mere computational trick and then
derived within the canonical equal-time (ET) quantization for the
Yang-Mills theories \cite{Bassettoetal1985}. Since then there were
attempts to derive the ML prescription for the LC gauge within the
LF canonical quantization \cite{McCartorRob}, \cite{Soldati},
however they were successful only for the free Abelian models with
two null surfaces used as the quantization surfaces. Now it becomes
clear that in order to have the ML-poles for the gauge propagators
in the interacting LF models one has to resort from the LC gauge
and choose another null axial gauge - the LF Weyl (LF Weyl)
condition: $A^{a}_{+} = A^{a-} = 0$. This gauge condition has
been introduced 
for the LF Quantum Electrodynamics (QED) in \cite{PrzNausKall}
within the DLCQ approach \cite{PauliBrodsky}. Quite recently it was
implemented for the perturbative QED in \cite{MoraraSoldati} and 
\cite{Przesz1998}, where the ML spurious poles have arised naturally
for the gauge field propagator. The fermionic currents in the QED
contain no derivatives of fields, thus the only complication which
appears in the LF quantization is connected with the presence of
nondynamical components of the fermion field $\psi_{-}$ and
$\psi_{-}^{\dag}$. This leads to the nonlocal (in $x^{-}$)
terms in the Hamiltonian, but ther consistent perturbation
theory is equivalent to the usual ET formulation.\footnote{There
are noncovariant contributions left, which are generated by the
1-dimensional (in $x^{-}$) determinants. Though these terms
formally depend on $A_{-}$ gauge field, then their
contributions can be neglected either on behalf of some proper
regularization chosen (the dimensional one \cite{MoraraSoldati}
or the Pauli-Villars one for fermions) or due to their
decoupling from normal modes of $A_{-}$ \cite{Przesz1998}.}
For the nonAbelian models things look worse, because the triple
gauge-gauge-gauge field couplings contain derivatives of fields and
for the LF Weyl gauge there will be also a time
derivative\footnote{For the LC gauge such time derivatives 
would be absent and this was the main reason for considering that
gauge condition in the early LF attempts \cite{Tomboulis73}, when
the CPV prescription was considered as consistent.}
$\partial_{+}A_i^a$ and the
canonical procedure would lead to the Dirac brackets which contain
interactions. We notice that a similar phenomenon also happens 
in the Abelian model of QED for charged scalar fields. For
a toy model in 1+1 
dimensions \cite{Przesz-hab} the consistent LF quantization has
been performed after a proper redefinition of scalar fields which
shifted all interaction into nonlocal (in $x^{-}$) interactions
but have left the LF Dirac brackets free of interactions. 
We may hope that a similar approach can be consistently applied
also for the nonAbelian model of SU(N) Yang-Mills fields. \\
Our present paper is organized as follows. In Section 2 we perform
the canonical quantization in three steps. First, we recapitulate
the analysis of the Abelian model for the helicity representation of 
the transverse components $A_i$. Second, we introduce the scalar
representation for the nonAbelian fields $A_i^a$ and choose a
suitable form of the LF Lagrangian. Third, the canonical LF
quantization is performed with the canonical free form of the LF
commutators. In Section 3 we define the generating functional for
all Green functions as the canonical phase-space path integral.
Within the perturbative theory, this path-integral is shown to be
equivalent to the canonical definition of the generating
functional, thus proving the consistency of the ML-prescription for
the LF Weyl gauge. In Section 4 these results are discussed and
further developments are sketched.

\section{Canonical quantization}
In this paper we implement the LF Weyl gauge condition
$A_{+}^a = 0$ explicitly for the SU(N) Yang-Mills theory, thus, 
the canonical Lagrangian, written in the LF notation, 
has the form 
\bey
{\cal L}_{YM}^{Weyl} & = & \frac 1 2 \left( \partial_{+}A_{-}^a
\right)^2 - \frac 1 4 \left(\partial_i A_j^a - D^{ab}_jA_i^b\right)^2
%\nonumber\\ & + & 
+ \partial_{+} A_i^a 
\left(D^{ab}_{-} A_i^b - \partial_{i}A_{-}^a\right) + A_\mu^a j^\mu_a,
\eey
where $D_\mu^{ab} = \partial_{\mu} \delta^{ab} + g f^{acb}
A^c_{\mu}$, $j^{\mu}_a$ describe arbitrary external sources and
the summation over repeating indices is understood. 
The transverse components of gauge fields can be
parameterized\footnote{Details of this parameterization and other 
notations are explained in the Appendix} by means of the
helicity fields $\phi^a_A, {\phi^{\dag}}^a_A$, which allow us
to write 
\bml 
\bey 
- \frac 1 4 \left(\partial_i A_j^a - D^{ab}_j A^a_i \right)^2 & = &
- \frac 1 4 \left( G^a_{AB} - 
G^a_{\bar{A} \bar{B}}\right) \left( G^a_{A\bar{B}} - G^a_{ \bar{A}
A}\right) - \frac 1 2 \left( G^a_{A\bar{B}}G^a_{\bar{A} B} + G^a_{AB}
G^a_{\bar{A} \bar{B}}\right),\\ 
\partial_{+} A_{i}^a \left(D^{ab}_{-} A_i^b - \partial_i A_{-}^a
\right) & = & \partial_{+} \phi^a_{A}\left( D_{-}^{ab}
{\phi^{\dag}}^a_A - \nabla_{\bar{A}} A^a_{-}\right) + \partial_{+}
{\phi^{\dag}}^a_{A}\left( D_{-}^{ab}\phi^b_A -  \nabla_{A} A^a_{-} \right).  
\eey \eml 
In this way we reach the expression for the Lagrangian density
\bey
{\cal L}_{YM}^{Weyl} & = & \frac 1 2 \left( \partial_{+}A_{-}^a
\right)^2 + \partial_{+} \phi^a_{A}\left( D_{-}^{ab}
{\phi^{\dag}}^a_A - \nabla_{\bar{A}} A^a_{-}\right) + \partial_{+}
{\phi^{\dag}}^a_{A}\left( D_{-}^{ab}\phi^b_A - \nabla_{A} A^a_{-}
\right)\nonumber\\ 
& - & \frac 1 4 \left( G^a_{AB} - 
G^a_{\bar{A} \bar{B}}\right) \left( G^a_{A\bar{B}} - G^a_{ \bar{A}
A}\right) - \frac 1 2 \left( G^a_{A\bar{B}}G^a_{\bar{A} B} + G^a_{AB}
G^a_{\bar{A} \bar{B}}\right)\nonumber\\
& + & A_{-}^a j^{-}_a
+ {\phi^{\dag}}^a_A s^a_A + {s^{\dag}}^a_A \phi^a_A,\label{primYMLagr}
\eey
where $\displaystyle s^{a}_A = \frac{j_{a}^{2A} - i j_a^{2A+1}}{\sqrt{2}},
\ {s^{\dag}}^{a}_A = \frac{j_{a}^{2A} +i j_a^{2A+1}}{\sqrt{2}}$.
However before starting the canonical quantization procedure we
integrate the second and third terms by parts in order to obtain
the equivalent Lagrangian density  
\bey
\widetilde{\cal L}_{YM}^{Weyl} & = & \frac 1 2 \left(
\partial_{+}A_{-}^a \right)^2 + \partial_{+} \phi^a_{A} D_{-}^{ab}
{\phi^{\dag}}^b_A
+ \partial_{+} {\phi^{\dag}}^a_{A} D_{-}^{ab}\phi^b_A
 - \partial_{+} A_{-}^a \left( \nabla_{\bar{A}}
\phi^a_{A} + \nabla_{A} {\phi^{\dag}}^a_{A} \right)\nonumber\\ 
& - & \frac 1 4 \left( G^a_{AB} - 
G^a_{\bar{A} \bar{B}}\right) \left( G^a_{A\bar{B}} - G^a_{ \bar{A}
A}\right) - \frac 1 2 \left( G^a_{A\bar{B}}G^a_{\bar{A} B} + G^a_{AB}
G^a_{\bar{A} \bar{B}}\right)\nonumber\\
& + & A_{-}^a j^{-}_a
+ {\phi^{\dag}}^a_A s^a_A + {s^{\dag}}^a_A \phi^a_A,\label{equiYMLagr}
\eey
which further will be taken as the starting point for quantization.
We stress here that one could also start with
(\ref{primYMLagr}) and then would reach the same final results,
however (\ref{equiYMLagr}) produces least possible
complications during the quantization procedure. Also we point
out that our formulation is performed within the dimensional
regularization ($D = 2\omega$), which at the stage of canonical
quantization allows for the proper definition of independent
modes, while at the level of perturbative calculation it
covariantly regularizes Feynman integrals.
 
\subsection{Canonical quantization for the Abelian case}

Before analysing the complete nonAbelian model, we will
present results for the Abelian case, when all color indices
are omitted and we put $g=0$ in (\ref{equiYMLagr}). Therefore
our first starting point is the equivalent Lagrangian density
for the Abelian case 
\bey
\widetilde{\cal L}_{Abel}^{Weyl} & = & \frac 1 2 \left( \partial_{+}A_{-}
\right)^2 + \partial_{+} \phi_{A} \partial_{-}\phi^{\dag}_A
+ \partial_{+} \phi^{\dag}_{A} \partial_{-}\phi_A
- \partial_{+} A_{-} \left( \nabla_{\bar{A}}
\phi_{A} + \nabla_{A} \phi^{\dag}_{A} \right)\nonumber\\ 
& - & \frac 1 4 \left( \nabla_A \phi_B - \nabla_B \phi_A
- \nabla_{\bar{A}} \phi^{\dag}_B + \nabla_{\bar{B}} \phi_A^{\dag}\right)
\left( \nabla_A \phi^{\dag}_B - \nabla_{\bar{B}} \phi_A
- \nabla_{B} \phi^{\dag}_A + \nabla_{\bar{A}} \phi_B\right)\nonumber\\
& - & \frac 1 2 \left( \nabla_A \phi^{\dag}_B - \nabla_{\bar{B}}
\phi_A\right)\left( \nabla_{\bar{A}} \phi_B - \nabla_{B}
\phi_A^{\dag} \right) 
- \frac 1 2 \left( \nabla_A \phi_B - \nabla_{B} \phi_A\right)
\left( \nabla_{\bar{A}} \phi^{\dag}_{B} - \nabla_{\bar{B}}
\phi_A^{\dag} \right) \nonumber\\
& + & A_{-} j^{-} + \phi^{\dag}_A s_A + s^{\dag}_A \phi_A
,\label{LagrAbel} 
\eey
and we easily find the relevant canonical momenta
\bml \bey
\Pi_{-} & = & \partial_{+} A_{-} - \nabla_{\bar{A}}\phi_A
- \nabla_{A} \phi^{\dag}_A,\\
\Pi_{\phi_A} & = & \partial_{-}\phi^{\dag}_A ,\\
\Pi_{\phi^{\dag}_A}  & = & \partial_{-}\phi_A, 
\eey \eml 
and the canonical Hamiltonian density\footnote{We have integrated
by parts the transverse partial derivatives in order to have the
most compact notation for the Hamiltonian density. Therefore the
sign $\approx$ means that some boundary terms, completely
superfluous in the forthcoming analysis, are omitted.}
\bey
{\cal H}_{can}^{Abel} & = & \Pi_{\phi_A} \partial_{+} \phi_A +
\Pi_{\phi^{\dag}_A} \partial_{+} \phi^{\dag}_A + \Pi_{-} \partial_{+} A_{-}
- {\cal L} \nonumber\\ 
& \approx &  \frac 1 2 \left(\Pi_{-} + \nabla_{\bar{A}}\phi_A 
+ \nabla_{A}\phi^{\dag}_A \right)^2 + \nabla_{B} \phi^{\dag}_A 
\nabla_{\bar{B}}\phi_A  + \nabla_{\bar{B}} \phi^{\dag}_A 
\nabla_{B}\phi_A  \nonumber\\
& - & \frac 1 2 \left( \nabla_{A} \phi^{\dag}_A 
+ \nabla_{\bar{A}}\phi_A \right)^2 
- A_{-} j^{-} - \phi^{\dag}_A s_A - s^{\dag}_A \phi_A.
\label{AbelcanHam}
\eey
The nonvanishing Dirac brackets are
\bml \label{AbelDirbra}\bey
2 \partial_{-}^x\left\{ \phi^{\dag}_A(x), \phi(y)_B\right\}_{x^{+}=
y^{+}} & =& - \delta_{AB} \delta^{2\omega -1}(\vec{x}- \vec{y}),\\
\left\{ A_{-}(x), \Pi_{-}(y)\right\}_{x^{+}=
y^{+}} & =& \delta^{2\omega -1}(\vec{x}- \vec{y}),
\eey \eml
and the Hamilton equations of motion 
\bml \bey
\partial_{+} \Pi_{-} & = & j^{-},\\
2\left(\partial_{+} \partial_{-} -
\nabla_A \nabla_{\bar{A}}  \right) \phi^{\dag}_B & = &
\nabla_B \Pi_{-} + s^{\dag}_B,\\
2\left(\partial_{+} \partial_{-} -
\nabla_A \nabla_{\bar{A}}  \right) \phi_B & = &
\nabla_{\bar{B}} \Pi_{-} + s_B,\\
\partial_{+} A_{-} & = & \Pi_{-} + \nabla_{\bar{A}}\phi_A
+ \nabla_{A} \phi^{\dag}_A
\eey \eml
are equivalent to the Euler-Lagrange equations which follow from
the Lagrangian density (\ref{LagrAbel}), thus proving the
consistency of the above canonical structure. \\ 
Next the canonical quantization follows directly; from
(\ref{AbelDirbra}) we obtain the commutation relations
\bml \label{Abelcancom}\bey
2 \partial_{-}^x\left[ \phi^{\dag}_A(x), \phi_B(y)
\right]_{x^{+} = y^{+}} & = & - i \delta_{AB} \delta^{2\omega -1}(\vec{x}-
\vec{y}),\\ 
\left[ A_{-}(x), \Pi(y)\right]_{x^{+}=
y^{+}} & =& i \delta^{2\omega -1}(\vec{x}- \vec{y}),
\eey \eml
while the quantum Hamiltonian is given exactly by
(\ref{AbelcanHam}). We see that source terms appear only linearly
thus the perturbative gauge field propagators will be given just by
the chronological products of respective (free) fields. However we
will not calculate these propagators here, instead we define the
path-integral representation for the Green functions as
\bey
Z[j^{-}, s_A^{\dag}, s_A] & = & {\cal N} \int {\cal D}A_{-} \ {\cal
D}\Pi_{-} \ {\cal D}\phi_A^{\dag} \ {\cal D}\phi_A \ \nonumber\\
&& \times \exp i \int d^{2\omega}x \ \left\{ \Pi_{-}
\partial_{+} A_{-} + \partial_{+}\phi^{\dag}_A \partial_{-} \phi_A
+ \partial_{-}\phi^{\dag}_A \partial_{+} \phi_A - {\cal
H}_{can}^{Abel}\right\}. 
\eey 
Because in the Hamiltonian $\Pi_{-}$ appears utmost quadratically,
then we can easily perform the Gaussian integral over $\Pi_{-}$
and obtain the desired result
\be
Z[j^{-}, s_A^{\dag}, s_A] = {\cal N} \int {\cal D}A_{-}  \ {\cal
D}\phi_A^{\dag} \ {\cal D}\phi_A \ \exp i \int d^{2\omega}x
\ {\cal L}_{Abel}^{Weyl}. 
\ee  
The remaining integrations are utmost Gaussian, thus one can
perform them rather easily and derive the explicit expression
for the generating functional
\bey
Z[j^{-}, s_A^{\dag}, s_A] & = & \exp{ - i \int d^{2\omega}x \
d^{2\omega} y \left[ s^{\dag}_A (x) D^{2\omega}_{F}(x-y) s_A(y)
+ j^{-}(x) \partial_{-}^x \Delta_{ML}^{2\omega}(x-y)
j^{-}(y)\right.}\nonumber\\ 
&& \left.- s^{\dag}_A (x) \nabla_{\bar{A}}^x \Delta_{ML}^{2\omega}
(x-y) j^{-}(y) - s_A(x) \nabla_A^x \Delta_{ML}^{2\omega}(x-y)
j^{-}(y)\right], 
\eey   
where
\bml \label{defsD_FDeltaML}\bey
D^{2\omega}_F(x) &= &i \int
\frac{d^{2\omega}k}{(2\pi)^4} \ \frac{e^{- i k \cdot x}}{2 
k_{+} k_{-} - k_\perp^2 + i \epsilon},\\ 
\Delta^{2\omega}_{ML}(x) & = & \int_0^{x^{+}}
 d\xi \ D_F^{2\omega}(\xi, \vec{x}) = - \int
\frac{d^{2\omega}k}{(2\pi)^4} \ \frac{e^{- i k \cdot x}}{2 
k_{+} k_{-} - k_\perp^2 + i \epsilon} \frac{1}{k_{+} + i
\epsilon' {\rm sgn}(k_{-})}.
\eey
\eml
The canonical quantum field propagators are defined as the
chronological products of operators and in the present case one
finds the following expressions 
\bml \bey
\left \langle 0 \left | T^{+} \phi_A(x) \phi_B^{\dag}(y)\right|
0 \right \rangle & = & \delta_{AB} D_F^{2\omega}(x-y), \\ 
\left \langle 0 \left | T^{+} \phi_A(x) A_{-}(y)\right| 0
\right \rangle & = & \nabla_{\bar{A}}^x
\Delta_{ML}^{2\omega}(x-y), \\ 
\left \langle 0 \left | T^{+} \phi^{\dag}_A(x) A_{-}(y)\right|
0 \right \rangle & = & \nabla_{A}^x
\Delta_{ML}^{2\omega}(x-y), \\ 
\left \langle 0 \left | T^{+} A_{-}(x) A_{-}(y)\right| 0 \right
\rangle & = & 2 \partial_{-}^x  \Delta_{ML}^{2\omega}(x-y), 
\eey \eml
therefore establishing the expected equivalence between the
canonical and path-integral definition of the generating functional
\be
Z[j^{-}, s^{\dag}_A, s_A] = \left \langle 0 \left | T^{+}
\exp{- i \int d^{2\omega}x \ \left( A_{-} j^{-} + \phi^{\dag}_A s_A
+ s^{\dag}_A \phi_A \right)}\right| 0 \right \rangle.
\ee

\subsection{Redefinition of transverse components of gauge fields}

Now we may start solve the main problem of this problem, namely, the
LF quantization of the nonAbelian gauge field system given by the
Lagrangian density (\ref{equiYMLagr}). Because the third and fourth
terms in (\ref{equiYMLagr}) look very similarly to those for the
charged scalar fields, therefore we will employ here the same trick
\cite{Przesz-hab}, \cite{Przesz99sca} which we have used in the
scalar QED. This means 
that we redefine charged complex fields $\phi_a$ and
$\phi_a^{\dag}$ as follows:
\bml
\bey
\phi^a_A& = & \chi^a_A + \sum_{n=1}^{\infty}
\left(\begin{array}{r} 
	-\frac 1 2 \\
n \end{array}\right)
\left(\widehat{a}\right)^n_{ab} \ast \chi_b = (\openone +
\widehat{a})^{-1/2}_{ab} \ast \chi_A^b,\\
{\phi^{\dag}}^a_A & = & {\chi^{\dag}}^a_A + {\chi^{\dag}}^b_A \ast 
\sum_{n=1}^{\infty} 
\left(\begin{array}{r} 
	-\frac 1 2 \\
n \end{array}\right)
\left({{\widehat{a}}^{\dag}}\right)^n_{ba}  =
{\chi^{\dag}}^b_A \ast (\openone + \widehat{a}^{\dag})^{-1/2}_{ba},\\ 
\widehat{a}_{ab} & = & g f^{acb}
\frac{1}{\partial_{-}} A^c_{-},\ \
\widehat{a}_{ab}^{\dag}  =  g f^{acb} A^c_{-}
\frac{1}{\partial_{-}} ,\\ 
\left(\widehat{a}\right)^n_{ab} & = & 
\left(\widehat{a}\right)^{n-1}_{ac} \ast 
\widehat{a}_{cb},
\eey
\eml
where the integration over $x^{-}$ is denoted by $\ast$. 
Next one can calculate the following expressions  
\bml
\bey
D^{ab}_{-}\phi^b_A & = & \left(1 +
\widehat{a}^{\dag}\right)^{1/2}_{ab} \ast 
\left(\partial_{-}\chi^b_A\right),\\ 
D^{ab}_{-}{\phi^{\dag}}^b_A & = &
\left(\partial_{-}{\chi^{\dag}}^a_A\right) \ast \left(1 +
\widehat{a}\right)^{1/2}_{ab}, \\ 
\partial_{\alpha} \phi^a_A & = & \left(\openone +
\widehat{a}\right)^{-1/2}_{ab} \ast
\left(\partial_{\alpha}\chi^b_A\right) +
L_{ab}[\partial_{\alpha}A_{-}] \ast \chi^b_A,\\ 
\partial_{\alpha} {\phi^{\dag}}^a_A & = & \left(\partial_{\alpha}
{\chi^{\dag}}^b_A \right) \ast \left(\openone +
\widehat{a}^{\dag}\right)^{-1/2}_{ba} + 
{\chi^{\dag}}^b_A \ast  R_{ba}[\partial_{\alpha}A_{-}] ,
\eey
where we have introduced the notation ($\partial_\alpha =
\partial_{+}, \nabla_{A}, \nabla_{\bar{A}}$) and 
\bey
L_{ab}[x^{-},y^{-}; \partial_\alpha A_{-}] & = & \int dz^{-} \
\partial_\alpha A_{-}^c (z^{-}) {\cal L}^{c}_{ab}[x^{-}, y^{-},
z^{-}],\\ 
R_{ab}[x^{-},y^{-}; \partial_\alpha A_{-}] & = & \int dz^{-} \
\partial_\alpha A_{-}^c (z^{-}) {\cal R}^{c}_{ab}[x^{-},y^{-},
z^{-},y^{-}] ,\\ 
{\cal L}_{ab}^c [x^{-}, y^{-}, z^{-}] & = & \frac{\delta}{\delta
A^c_{-}(z^{-})} \left(\openone + \widehat{a}\right)^{-1/2}_{ab}
[x^{-}, y^{-}],\\ 
{\cal R}_{ab}^c [x^{-}, y^{-}, z^{-}] & = & \frac{\delta}{\delta
A^c_{-}(z^{-})} \left(\openone + \widehat{a}^{\dag}
\right)^{-1/2}_{ab} [x^{-}, y^{-}]. 
\eey
\eml
The transverse components of the gauge field strength can be
expressed in terms of these new fields $\chi_A^a$ and
${\chi^{\dag}}^a_A$, for example 
\bml
\bey
G^a_{AB} & = & \left( \openone + \widehat{a} \right)^{-1/2}_{ab}
\ast \nabla_A \chi^b_B -  \left( \openone + \widehat{a} \right)^{-1/2}_{ab}
\ast \nabla_B \chi^b_A + L_{ab}[\nabla_{A} A_{-}] \ast \chi^b_B
\nonumber\\ 
& - & L_{ab}[\nabla_{B} A_{-}] \ast \chi^b_A
+ gf^{abc} \left( \openone + \widehat{a} \right)^{-1/2}_{bd}
\ast \chi^d_A \ \left( \openone + \widehat{a} \right)^{-1/2}_{ce}
\ast \chi^e_B = g_{AB}^a,\\
G^a_{\bar{A}B} & = & \left( \openone + \widehat{a} \right)^{-1/2}_{ab}
\ast \nabla_{\bar{A}} \chi^b_B - \nabla_B {\chi^{\dag}}^b_A \ast \left(
\openone + \widehat{a}^{\dag} \right)^{-1/2}_{ba} +
L_{ab}[\nabla_{\bar{A}} A_{-}] \ast \chi^b_B 
\nonumber\\ 
& - & {\chi^{\dag}}^b_A \ast R_{ab}[\nabla_{B} A_{-}] 
+ gf^{abc} {\chi^{\dag}}^d_A \ast \left( \openone + \widehat{a}^{\dag}
\right)^{-1/2}_{db} \  \left( \openone + \widehat{a} \right)^{-1/2}_{ce}
\ast \chi^e_B = g_{\bar{A}B}^a,
\eey
\eml
and the similar expressions for $g_{A\bar{A}}^a = G_{A\bar{A}}^a$
and $g_{\bar{A} \bar{B}}^a= G_{\bar{A} \bar{B}}^a$. All above
formulas allow us to write the redefined Lagrangian density 
as follows
\bey
{\cal L}_{mod} & = & \partial_{+} {\chi^{\dag}}^a_A \partial_{-}
\chi^{a}_A + \partial_{-} {\chi^{\dag}}^a_A \ast \partial_{+}
\chi^{a}_A +  \frac 1 2 \left( \partial_{+}A_{-}^a \right)^2  
+ \partial_{+} A_{-}^a {\cal J}^a - {\cal V} \nonumber\\
&& + A^a_{-} j^{-}_a + {s^{\dag}}^{a}_A \left(\openone +
\widehat{a} \right)^{1/2}_{ab} \ast \chi^{b}_A + {\chi^{\dag}}^a_A
\ast \left(\openone + \widehat{a}^{\dag} \right)^{1/2}_{ab}
s^b_A,\label{YMmodLagr} 
\eey 
where for brevity we have introduced the notation 
\bml \bey
{\cal J}^a & = & {\chi^{\dag}}^b_A \ast {\cal R}^{a}_{bc} \ast
\left(\openone + \widehat{a}^{\dag}\right)^{1/2}_{cd} \ast
\chi^d_A + {\chi^{\dag}}^b_A \ast \left(\openone +
\widehat{a}\right)^{1/2}_{bc} \ast {\cal L}^{a}_{cd} \ast 
\chi^d_A - {\chi^{\dag}}^b_A \ast {R}_{ba}[\nabla_{A}A_{-}] -
\nabla_{A}{\chi^{\dag}}^b_{A} 
\ast \left(\openone + \widehat{a}^{\dag}\right)^{-1/2}_{ba}
  \nonumber\\ 
&& -  {L}_{ab}[\nabla_{\bar{A}}A_{-}] \ast {\chi^{a}}_A -
\left(\openone + \widehat{a}\right)^{-1/2}_{ab} \ast 
\nabla_{\bar{A}}\chi^b_A , \\
{\cal V} & = & \frac 1 4 \left( g^a_{AB} - 
g^a_{\bar{A} \bar{B}}\right) \left( g^a_{A\bar{B}} - g^a_{ \bar{A}
A}\right) + \frac 1 2 \left( g^a_{A\bar{B}}g^a_{\bar{A} B} +
g^a_{AB} g^a_{\bar{A} \bar{B}}\right)
\eey
\eml 
and also we have used the identities
\be
\left(\openone + \widehat{a}^{\dag}\right)^{-1/2}_{ab} \ast \left(
\openone + \widehat{a}^{\dag}\right)^{1/2}_{bc} = 
\left(\openone + \widehat{a}\right)^{1/2}_{ab} \ast \left(\openone
+ \widehat{a}\right)^{-1/2}_{bc} = \delta_{ac}.
\ee
Now we are ready to start the canonical procedure and we find
the canonical conjugated momenta to $\chi^a_A$, ${\chi^{\dag}}^a_A$ and
$A_{-}^a$ respectively
\bml \bey
\Pi_{{\chi}^a_A} & = & \partial_{-} {\chi^{\dag}}^a_A\\
\Pi_{{\chi^{\dag}}^a_A} & = & \partial_{-} \chi^a_A\\
\Pi_{-}^a & = & \partial_{+} A_{-}^a + {\cal J}^a.
\eey \eml
Thus the nondynamical momenta conjugated to transverse fields
$\chi^a_A$ and ${\chi^{\dag}}^a_A$ have the free form and we
expected that they will not generate the interaction dependent
Dirac brackets. The price of this success is the more
complicated structure of interaction terms but this is not a
fundamental difficulty both in the canonical quantization procedure
and the later perturbation calculations. One easily can find
the canonical Hamiltonian density 
\bey 
{\cal H}_{can} & = & \Pi_{{\chi}^a_A} \ \partial_{+}\chi^a_A +
\Pi_{{\chi^{\dag}}^a_A}\ \partial_{+}{\chi^{\dag}}^a_A +
\Pi^{a}_{-}\ast \ \partial_{+} A_{-} - {\cal L} \nonumber\\
& = & \frac 1 2 \left( \Pi_{-}^{a} - {\cal J}^a\right)^2
+ {\cal V} -  {s^{\dag}}^{a}_A  \left(\openone + \widehat{a}^{\dag}
\right)^{-1/2}_{ab} \ast \chi^{b}_A - {\chi^{\dag}}^a_A \ast \left(
\openone + \widehat{a}^{\dag} \right)^{-1/2}_{ab} s^b_A -
A^a_{-}j^{-}_a, 
\label{YMcanHam}
\eey 
and then the canonical commutation rules which are direct
generalization of those (\ref{Abelcancom}) in the Abelian case 
\bml \label{nonAbelcancom}\bey
2 \partial_{-}^x\left[ {\phi^{\dag}}^a_A(x), \phi(y)^b_B\right]_{x^{+}=
y^{+}} & =& - i\delta^{ab} \ \delta_{AB} \ \delta^{2\omega -1}(\vec{x}- \vec{y}),\\
\left[ A_{-}^a(x), \Pi_{-}^b(y)\right]_{x^{+}=
y^{+}} & =& i \delta^{ab} \ \delta^{2\omega -1}(\vec{x}- \vec{y}).
\eey \eml
One can check that these commutators and the Hamiltonian
(\ref{YMcanHam}) generate Heisenberg equations of motion 
\bml \bey
2 \partial_{+} \partial_{-} \chi^a_A & = & \partial_{+} A_{-}^b \ast
\frac{\delta {\cal J}^b}{\delta {\chi^{\dag}}_A^a} - \frac{\delta {\cal
V}}{\delta {\chi^{\dag}}^a_A} + \left(\openone + \widehat{a}^{\dag}
\right)^{-1/2}_{ab}\ast s^b_A, \\
2 \partial_{+} \partial_{-} {\chi^{\dag}}^a_A & = & \partial_{+}
A_{-}^b \ast \frac{\delta {\cal J}^b}{\delta \chi^a_A} -
\frac{\delta {\cal V}}{\delta \chi^a_A} + {s^{\dag}}^b_A \ast \left(1 +
\widehat{a}\right)^{-1/2}_{ba},\\
\partial_{+} \Pi^a_{-} & = & \left(\Pi^b_{-} - {\cal J}^b \right)
\ast \frac{\delta {\cal J}^b}{\delta A_{-}^a} - \frac{\delta {\cal
V}}{\delta A_{-}^a} + j^{-}_a + {s^{\dag}}^b_A \ast {\cal L}^a_{bc} \ast
\chi^c_A +  {\chi^{\dag}}^b_A \ast {\cal R}^a_{bc} \ast s^c_A,\\
\partial_{+} A^a_{-} & = & \Pi^a_{-} - {\cal J}^a,
\eey \eml
which, modulo proper ordering of non-commuting terms, are
equivalent to the Euler-Lagrange equations generated from the
Lagrangian (\ref{YMmodLagr}). 

\section{Generating functional}

Having found the canonical structure for the gauge system we can
define the generating functional for Green functions as the
path-integral over phase space
\bey
Z[s^a_A,{s^{\dag}}^a_A, j^{-}_a] &\stackrel{df}{=} & \int {\cal
D}A_{-}^a \ {\cal D}\Pi^{-}_a \ {\cal D} \chi^a_A \ {\cal D}
{\chi^{\dag}}^a_A \nonumber\\ 
&& \ \ \ \exp{i \int d^{2\omega}x \ \left[2\partial_{-}{\chi^{\dag}}^a_A 
\partial_{+}\chi^a_A + 2 \partial_{+}{\chi^{\dag}}^a_A
\partial_{-} \chi^a_A  
+ \Pi^{a}_{-} \partial_{+} A_{-}^a - H_{can}\right]}.
\eey
Next we can perform the Gaussian integration over $\Pi_{-}^a$ and
rewrite the generating functional 
\bey
Z[s^a_A,{s^{\dag}}^a_A, j^{-}_a] &\stackrel{df}{=} & \int {\cal
D}A_{-}^a \ {\cal D} \chi^a_A \ {\cal D} {\chi^{\dag}}^a_A \ \exp{i
\int d^{2\omega}x {\cal L}_{mod}}. 
\eey
Then we may reinstall linear couplings to the external sources
$s^a_A$ and ${s^{\dag}}^a_A$ via the following change of path variables
\bml \bey
\chi^a_A & = & \left(1 + \widehat{a}\right)^{1/2}_{ab} \ast
\phi^b_A,\\ 
{\chi^{\dag}}^a_A & = & {\phi^{\dag}}^b_A \ast \left(1 +
\widehat{a}^{\dag}\right)^{1/2}_{ba},
\eey \eml 
which gives the main result of this paper:
\bey
Z[s^a_A,{s^{\dag}}^a_A, j^{-}_a] &\stackrel{df}{=} & \int {\cal
D}A_{-}^a \ {\cal D} \phi^a_A \ {\cal D}{\phi^{\dag}}^a_A \
Det^{1/2}\left(1 + \widehat{a}\right) \  Det^{1/2} \left(1 +
\widehat{a}^{\dag}\right) \ \exp{i \int d^{2\omega}x {\cal L}_{YM}^{Weyl}}.
\eey
We notice, after \cite{MoraraSoldati}, that within the dimensional
regularization we can safely omit the functional determinants as
irrelevant constants. Therefore our LF quantization leads to the
same path-integral definition of generating functional as the
commonly used ET formulation.

In the Abelian case we have checked that the generating functional
can be equivalently defined either as the path-integral or
canonically as the chronological product. Below we will check the
respectful equivalence for the Yang-Mills case and we start with
the canonical definition 
\be
Z[j_a^{-}, {s^{\dag}}^a_A, s^a_A] = \left \langle 0 \left | T^{+}
\exp{- i \int d^{2\omega}x \ \left({\cal H}_{int}[A_{-}^a, \Pi^a_{-},
{\chi^{\dag}}^a_A, \chi^a_A, {s^{\dag}}^a_A, s^a_A] - A^a_{-} j_a^{-}
\right)}\right| 0 \right \rangle. 
\ee
where
\bml \bey
{\cal H}_{int}[A_{-}^a, \Pi^a_{-}, {\chi^{\dag}}^a_A, \chi^a_A,
{s^{\dag}}^a_A, s^a_A] & = & {\cal H}_{can} - {\cal H}_{0}\nonumber\\
& = & \frac{1}{2} \left( {\cal J}^a_{int}\right)^2 + 
{\cal V}_{int} - {\cal J}^a_{int} \left( \Pi^a_{-} - {\cal
J}^a_{0} \right) - {\chi^{\dag}}^a_A \ast \left(1 + 
\widehat{a}^{\dag} \right)^{-1/2}_{ab}  s^b_A,\\ 
{\cal H}_{0} & = & \frac{1}{2} \left( \Pi^a_{-} - {\cal
J}^a_0\right)^2 + {\cal V}_{0}^a,\\
{\cal J}^a_0 & = & \nabla_A {\chi^{\dag}}^a_A + \nabla_{\bar{A}}
{\chi}^a_A,\\ 
{\cal V}_0 & = & \nabla_B {\chi^{\dag}}^a_A \nabla_{\bar{B}}
{\chi}^a_A + \nabla_{\bar{B}} {\chi^{\dag}}^a_A \nabla_{B}
{\chi}^a_A - \frac 1 2 \left( \nabla_A {\chi^{\dag}}^a_A +
\nabla_{\bar{A}} {\chi}^a_A \right)^2,\\
{\cal J}^a_{int} & = & {\cal J}^a - {\cal J}^a_0,\\
{\cal V}_{int} & = & {\cal V} - {\cal V}_0.
\eey \eml
Introducing auxiliary sources for the canonical fields $\Pi^a_{-}$,
$\chi^a_A$ and ${\chi^{\dag}}^a_A$ we can take the modified external
Hamiltonian density 
\be
{\cal H}_{ext} = - \Pi^a_{-} k_a - {\chi^{\dag}}^a_A p^a_A -
{p^{\dag}}^a_A \chi^a_A - A_{-}^a j^{-}_a 
\ee
and this allows to factorize the interaction Hamiltonian outside
the vacuum expectation value 
\be
Z[j_a^{-}, {s^{\dag}}^a_A, s^a_A] = \left.\exp{ - i\int d^{2\omega}x \ {\cal
H}_{int}\left[\widetilde{A}_{-}^a, \widetilde{\Pi}^a_{-},
{\widetilde{\chi^{\dag}}}^a_A, {\widetilde{\chi}}^a_A, {s^{\dag}}^a_A, 
s^a_A \right]} \left \langle 0 \left | T^{+} 
\exp{- i \int d^{2\omega}x \ {\cal H}_{ext} }\right| 0 \right
\rangle\right|_{k = p = p^{\dag} = 0}, 
\ee
where ${\displaystyle \widetilde{A}}_{-}^a  = 
\frac{i\delta}{\delta j^{-}_a}, \  
\widetilde{\Pi}_{-}^a  =  \frac{i\delta}{\delta k_a}, \
{\widetilde{\chi}}^a_A  =  \frac{i\delta}{\delta {p^{\dag}}^a_A}, \
{\widetilde{\chi^{\dag}}}^a_A  =  \frac{i\delta}{\delta p^a_A}.$ 
First we analyse the free generating functional 
\bey
Z_0[j_a^{-}, k^a, {p^{\dag}}^a_A, p^a_A] & = & \left \langle 0 \left |
T^{+} \exp{- i \int d^{2\omega}x \ {\cal H}_{ext} }\right| 0 \right \rangle
\nonumber\\
 & = & \exp{ - i \int d^{2\omega}x \
d^{2\omega} y \left[ p^{\dag}_a (x) (x) D^{2\omega}_{F}(x-y) p_a(y)
+ p^{\dag}_a(x) \partial_{-}^x \Delta_{ML}^{2\omega}(x-y)
j_b^{-}(y) \right.}\nonumber\\ 
&& \left.- p_a(x)_A (x) \nabla_{\bar{A}}^x \Delta_{ML}^{2\omega}
(x-y) j^{-}_a(y) - {p^{\dag}}_A^a(x) \nabla_A^x \Delta_{ML}^{2\omega}(x-y)
j^{-}_a(y) + k^{\dag}_a (x) E^1_F(x-y) j^{-}_a(y)\right], 
\eey
where $D_F(x)$ and $\Delta_{ML}(x)$ are given by Eqs.
(\ref{defsD_FDeltaML}), 
while $E_F^1(x)$ is defined as
\be
E_F^1(x)  =    \int \frac{d^{2\omega}k}{(2\pi)^{2\omega}}
\frac{e^{-i k\cdot (x-y)}}{k_{+} + i \epsilon {\rm
sgn}(k_{-})}, 
\ee
Thus the free generating functional has its path-integral
representation 
\bey
Z_0[j_a^{-}, k^a_A, {p^{\dag}}_A^a, p^a_A] & = & {\cal N} \int
{\cal D}A_{-}^a \ 
{\cal D} \chi_A^a \ {\cal D} {\chi^{\dag}}^a_A {\cal D}\Pi_{-}^a
\nonumber\\ 
&& \times
\exp{ i \int d^{2\omega}x \ \left(\partial_{+}{\chi^{\dag}}^a_A
\partial_{-} \chi^a_A + \partial_{-}{\chi^{\dag}}^a_A
\partial_{+} \chi^a_A + \partial_{+}A_{-}^a \Pi_{-}^a - {\cal
H}_{0} - {\cal H}_{ext} \right) }
\eey
In the complete generating functional, the exponential operator can
be pushed under the sign of path integration and then easily one
obtains the expected result
\bey
Z[s^a_A,{s^{\dag}}^a_A, j_a^{-}] & = & {\cal N}' \int {\cal D}A_{-}^a \
{\cal D} \chi_A^a \ {\cal D} {\chi^{\dag}}^a_A {\cal D}\Pi_{-}^a
\nonumber\\ 
&& \times
\exp{ i \int d^{2\omega}x \ \left(\partial_{+}{\chi^{\dag}}^a_A
\partial_{-} \chi^a_A + \partial_{-}{\chi^{\dag}}^a_A
\partial_{+} \chi^a_A + \partial_{+}A_{-}^a \Pi_{-}^a - {\cal
H}_{can} \right) }. 
\eey

\section{Conclusions}
In this paper we have shown how the nonAbelian couplings can be
consistently quantized within the LF approach without spoiling the
natural ML-prescription for the spurious poles. While we have used
only formal arguments for checking the equivalence of this approach
with the usual ET quantization, it will be interesting to check how
it works for the explicit calculations. We think that the
computation of the gauge invariant quantities like the Wilson loops
will be a quite interesting cross-check. Further we expect that also
other choices of gauge conditions can be similarly implemented for
the LF Yang-Mills theory\footnote{This would be a generalization of
the fermionic QED analysis from \cite{Przesz-hab}.}

\appendix
\section{Notations}
In the space-time of $2\omega$ dimensions we introduce the
light-front notation: 2 longitudinal coordinates $\displaystyle
x^{\pm} = \frac{x^{0} \pm x^{2\omega -1}}{\sqrt{2}}$ and $2(\omega
-1)$ transverse coordinates $x^i$. Similarly we denote components
of any vector $V_\mu$: $\displaystyle V_{\pm} = \frac{V_{0} \pm
V_{2\omega -1}}{\sqrt{2}}, V_i$. Further, the transverse components
of the gauge field potential $A_i^a, \ i = 1, \dots, 2(\omega -1)$
are parameterized by the pair of Hermitian fields
$\left(\phi_A^a, {\phi^{\dag}}_A^a, \ A=1, \ldots, \omega
-1\right)$ as follows: 
\bml
\bey
A_{2A}^a & =& \frac{\phi_A^a + {\phi^{\dag}}_A^a}{\sqrt{2}},\\
A_{2A+1}^a & =& i \frac{\phi_A^a - {\phi^{\dag}}_A^a}{\sqrt{2}}.
\eey
\eml
When this notation is implemented for into the components of the gauge
field $F^a_{ij}$ we obtain
\bml
\bey
F^a_{2A \ 2B} & = & \partial_{2A} A^a_{2B} - \partial_{2B} A^a_{2A}
+ g f^{abc} A^b_{2A} A^c_{2B} = \frac 1 2 \left ( G^a_{AB} + 
G^a_{\bar{A}B} + G^a_{A\bar{B}} + G^a_{\bar{A}\bar{B}}\right),\\
F^a_{2A \ 2B+1} & = & \partial_{2A} A^a_{2B+1} - \partial_{2B+1}
A^a_{2A} + g f^{abc} A^b_{2A} A^c_{2B+1} = \frac i 2 \left (
G^a_{AB} + G^a_{\bar{A}B} - G^a_{A\bar{B}} -
G^a_{\bar{A}\bar{B}}\right),\\ 
F^a_{2A+1 \ 2B+1} & = & \partial_{2A+1} A^a_{2B+1} -
\partial_{2B+1} A^a_{2A+1} + g f^{abc} A^b_{2A+1} A^c_{2B+1} =
\frac 1 2 \left ( - G^a_{AB} + 
G^a_{\bar{A}B} + G^a_{A\bar{B}} - G^a_{\bar{A}\bar{B}}\right),
\eey 
\eml
where
\bml \bey
G^a_{A B} & = & \nabla_{A} \phi^a_{B} - \nabla_{B} \phi^a_{A}
+ g f^{abc} \phi^b_{A} \phi^c_{B},\\
G^a_{\bar{A} B} & = & \nabla_{\bar{A}} \phi^a_{B} - \nabla_{B}
{\phi^{\dag}}^a_{A} + g f^{abc} {\phi^{\dag}}^b_{A} \phi^c_{B},\\
G^a_{{A} \bar{B}} & = & \nabla_{A} {\phi^{\dag}}^a_{B} -
\nabla_{\bar{B}} {\phi}^a_{A} + g f^{abc} {\phi}^b_{A}
{\phi^{\dag}}^c_{B} = \left(G^a_{\bar{A}B}\right)^{\dag},\\ 
G^a_{\bar{A}\bar{B}} & = & \nabla_{\bar{A}} {\phi^{\dag}}^a_{B} -
\nabla_{\bar{B}} {\phi^{\dag}}^a_{A} + g f^{abc} {\phi^{\dag}}^b_{A}
{\phi^{\dag}}^c_{B} = \left(G^a_{AB}\right)^{\dag},\\
\nabla_{A} & = & \frac{\partial_{2A} - i
\partial_{2A+1}}{\sqrt{2}}, \ \ \nabla_{\bar{A}} =
\frac{\partial_{2A} + i \partial_{2A+1}}{\sqrt{2}}.
\eey \eml
The above notation allows us to write the following useful
relations which are used in the main text
\bml
\bey
\sum_{i,j = 1}^{2(\omega -1)} \left( F^a_{ij}\right)^2 & = & 
\sum_{A,B = 1}^{\omega -1} \left[ \left( G^a_{AB} -
G^a_{\bar{A} \bar{B}}\right) \left( G^a_{A\bar{B}} -
G^a_{ \bar{A} A}\right) + 2 \left( G^a_{A\bar{B}}G^a_{\bar{A} B}
+  G^a_{AB} G^a_{\bar{A} \bar{B}}\right) \right],\\
\sum_{i=1}^{2(\omega -1)} \partial_{+} A_{i}^a \partial_i
A_{-}^a & = & \sum_{A = 1}^{\omega -1} \left[ \partial_{+}
\phi^a_{A} \nabla_{\bar{A}} A^a_{-} + \partial_{+}
{\phi^{\dag}}^a_{A} \nabla_{A} A^a_{-} \right],\\
\sum_{i=1}^{2(\omega -1)} \partial_{+} A_{i}^a D^{ab}A_{i}^a & = &
\sum_{A = 1}^{\omega -1} \left[ \partial_{+} 
\phi^a_{A} D_{-}^{ab} {\phi^{\dag}}^b_{A} + \partial_{+}
{\phi^{\dag}}^a_{A} D_{-}^{ab} \phi^b_{A} \right]\\
\sum_{i,j=1}^{2(\omega -1)} \partial_{i} A_{j} \partial_i A_{j} & = &
\sum_{A,B = 1}^{\omega -1} 2 \left[ \nabla_B \phi^{\dag}_{A}
\nabla_{\bar{B}} \phi_{A} + \nabla_{\bar{B}} \phi^{\dag}_{A}
\nabla_{B} \phi_{A} \right], \\
\sum_{i=1}^{2(\omega -1)} \partial_{i} A_{i} & = &
\sum_{A = 1}^{\omega -1}  \left[ \nabla_A \phi^{\dag}_{A}
 + \nabla_{\bar{A}} \phi_{A}\right].
\eey
\eml
In the main text we tacitly use the convention of summing over
repeating indices over all possible values of different
indices, thus in the above formulas we omit the sum signs
hoping this will lead to no confusion or difficulty.

\end{document}